\documentclass[twocolumn, superscriptaddress, aps]{revtex4-2}

\usepackage{amsmath}
\usepackage{amssymb}
\usepackage{wasysym}
\usepackage{graphicx}
\usepackage{hyperref}
\usepackage{color}
\usepackage{physics}
\usepackage{siunitx}
\usepackage{xcolor}
\usepackage{changes}
\usepackage{comment}
\usepackage{bm}
\usepackage{natbib}
\usepackage[utf8]{inputenc}
\usepackage{tikz}
\usepackage{transparent}
\usepackage{braket}
\usepackage{commath}

\usepackage{titlesec}
\titlespacing*{\subsection}{0pt}{1\baselineskip}{0.5\baselineskip}

\allowdisplaybreaks

\begin{document}

\title{Evaporation of microwave-shielded polar molecules to quantum degeneracy}

\author{Andreas Schindewolf}
\author{Roman Bause}
\author{Xing-Yan Chen}
\author{Marcel Duda}
\affiliation{Max-Planck-Institut f\"{u}r Quantenoptik, 85748 Garching, Germany}
\affiliation{Munich Center for Quantum Science and Technology, 80799 M\"{u}nchen, Germany}
\author{Tijs Karman}
\affiliation{Institute for Molecules and Materials, Radboud University, Heijendaalseweg 135, 6525 AJ Nijmegen, Netherlands}
\author{Immanuel~Bloch}
\affiliation{Max-Planck-Institut f\"{u}r Quantenoptik, 85748 Garching, Germany}
\affiliation{Munich Center for Quantum Science and Technology, 80799 M\"{u}nchen, Germany}
\affiliation{Fakult\"{a}t f\"{u}r Physik, Ludwig-Maximilians-Universit\"{a}t, 80799 M\"{u}nchen, Germany}
\author{Xin-Yu~Luo} \email{e-mail: xinyu.luo@mpq.mpg.de}
\affiliation{Max-Planck-Institut f\"{u}r Quantenoptik, 85748 Garching, Germany}
\affiliation{Munich Center for Quantum Science and Technology, 80799 M\"{u}nchen, Germany}

\date{\today}

\begin{abstract}
Ultracold polar molecules offer strong electric dipole moments and rich internal structure, which makes them ideal building blocks to explore exotic quantum matter \cite{Baranov2002,Cooper2009,Shi2010,Wu2016,Schmidt2021,Bruun2008,Matveeva2012,Baranov2012}, implement novel quantum information schemes \cite{DeMille2002,Yelin2006,Ni2018}, or test fundamental symmetries of nature
\cite{Safronova2018}. Realizing their full potential requires cooling interacting molecular gases deeply into the quantum degenerate regime. However, the complexity of molecules which makes their collisions intrinsically unstable at the short range, even for nonreactive molecules \cite{Mayle2013,Christianen2019,Gregory2020,Liu2020,Bause2021a,Gersema2021}, has so far prevented the cooling to quantum degeneracy in three dimensions. Here, we demonstrate evaporative cooling of a three-dimensional gas of fermionic sodium-potassium molecules to well below the Fermi temperature using microwave shielding. The molecules are protected from reaching short range with a repulsive barrier engineered by coupling rotational states with a blue-detuned circularly polarized microwave. The microwave dressing induces strong tunable dipolar interactions between the molecules, leading to high elastic collision rates that can exceed the inelastic ones by at least a factor of $460$. This large elastic-to-inelastic collision ratio allows us to cool the molecular gas down to $21$ nanokelvin, corresponding to $0.36$ times the Fermi temperature. Such unprecedentedly cold and dense samples of polar molecules open the path to the exploration of novel many-body phenomena, such as the long-sought topological $p$-wave superfluid states of ultracold matter.
\end{abstract}

\maketitle

The field of ultracold polar molecules is currently on the verge of entering an exciting phase of showing its full potential in quantum sciences \cite{Lincoln2009,bohn2017}. The tunable anisotropic long-range interactions and rich internal degrees of freedom of polar molecules open up the possibility to study exotic quantum phases ranging from $p$-wave superfluids \cite{Baranov2002,Cooper2009,Shi2010}, supersolids \cite{Wu2016,Schmidt2021}, and 
Wigner crystals \cite{Bruun2008,Matveeva2012} to novel spin systems and extended Hubbard models \cite{Baranov2012} and can also be utilized for quantum computation \cite{DeMille2002,Yelin2006,Ni2018}. Many of these proposals require a deeply degenerate quantum gas of molecules at tens of nanokelvin with strong dipolar interactions in three dimensions (3D).

While non-interacting polar molecules can partially inherit low entropy from degenerate atomic mixtures \cite{Moses2015,DeMarco2018,Duda2021}, active cooling to the quantum degenerate regime has remained challenging. The main hurdle towards efficient evaporative cooling of ultracold molecules is posed by their rapid collisional loss. Especially when the molecules are polarized by an electric field in 3D traps they tend to collapse in attractive head-to-tail collisions \cite{Ni2010dipolar}. It turned out that even molecules that are nominally stable against chemical reactions \cite{Zuchowski2010reactions} undergo inelastic collisions when they reach short range. The exact nature of these loss processes is still not fully understood and remains under investigation \cite{Mayle2013,Christianen2019,Gregory2020,Liu2020,Bause2021a,Gersema2021}.

Collisional loss in short-range encounters between molecules can be suppressed by engineering repulsive interactions using external fields \cite{Avdeenkov2006, Gorshkov2008,Micheli2010,Gonzalez2017,Matsuda2020}. 
Recently, a molecular gas of $^{40}$K$^{87}$Rb was stabilized and evaporatively cooled to below the Fermi temperature by applying a strong dc electric field and confining the motion of the molecules to two dimensions (2D), which effectively prevented attractive collisions along the third dimension \cite{Valtolina2020}. For molecules in 3D, reduced collisional loss has been demonstrated by using a specific dc electric field \cite{Li2021} to bring rotational states of colliding molecules in resonance with each other. However, attempts to produce a degenerate quantum gas of polar molecules via collisional cooling so far remained unsuccessful in 3D due to a low elastic-to-inelastic collision ratio as well as a relatively low initial phase space density \cite{Son2020,Li2021}. Repulsive barriers between molecules can also be engineered by applying blue-detuned circularly polarized microwaves \cite{Karman2018}, which was successfully employed recently to reduce collisional loss between two CaF molecules in an optical tweezer \cite{Anderegg2021}.

In this work, we induce fast elastic dipolar collisions in a quantum gas of fermionic $^{23}$Na$^{40}$K molecules dressed by a circularly polarized microwave field while strongly suppressing inelastic collisions in all three dimensions. The elastic collision rate increases dramatically with the effective lab-frame dipole moment, which can be adjusted with the microwave power and detuning. Under appropriate conditions, we find that the elastic collision rate can be about $500$ times larger than the inelastic collision rate, and even exceed the trap frequencies such that the molecular gas enters the hydrodynamic regime. We evaporate a 3D near-degenerate sample of molecules down to $21(5)$\,nK, which corresponds to $36(9)$\% of the Fermi temperature $T_\text{F}$, deep in the quantum degenerate regime. At such low temperatures, the inelastic collision rate between molecules becomes negligible, leading to a long lifetime of the degenerate molecular sample of up to $0.6$\,s, mainly limited by residual one-body loss induced by the technical noise of the microwave.

\section*{Microwave shielding and dipolar elastic collisions}
To realize shielding in 3D, the two lowest rotational states of the molecules can be coupled via a blue-detuned circularly polarized microwave \cite{Karman2018}. The mixed state features an induced rotating dipole moment that follows the strong ac electric field of the microwave. Molecules prepared in this state exhibit an effective dipolar interaction at long range. When they approach each other, the direct coupling between the induced dipoles starts to dominate, forming a repulsive barrier that effectively shields the molecules from detrimental inelastic collisions at short range.

\begin{figure}
\centering
\includegraphics[width = \linewidth]{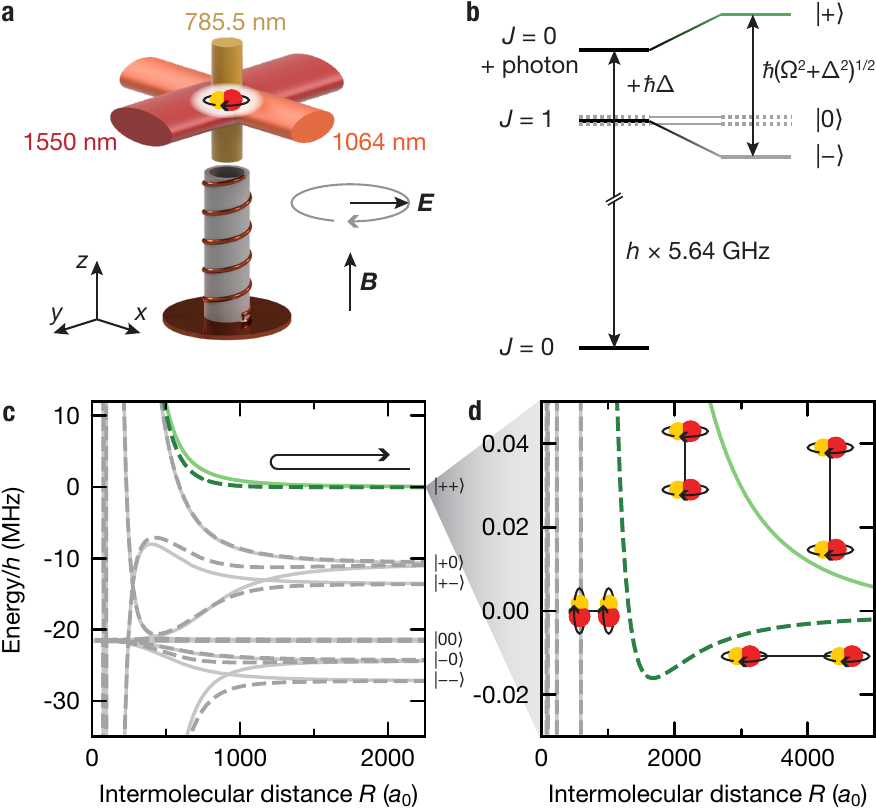}
\caption{\textbf{Microwave shielding.} \textbf{a}, Sketch of the experimental setup. The molecules are confined by up to three optical dipole traps with the indicated wavelengths. A helical antenna emits the rotating electric field $\vec{E}$. A dc magnetic offset field $\vec{B}$ is generated by a pair of coils (not shown). \textbf{b}, Preparation of the molecules in the dressed state $|{+}\rangle$. The microwave detuning $\Delta$ is typically comparable with the Rabi frequency $\Omega$. \textbf{c}, Interaction potentials for $\Delta = 2\pi \times 8$\,MHz and $\Omega = 2\pi \times 11$\,MHz. The solid and dashed lines show the potential energy of molecules colliding along ($\theta = 0$) and orthogonal ($\theta = 90^\circ$) to the microwave propagation direction, respectively. The centrifugal barrier is omitted for clarification. Here, $a_0$ is the Bohr radius. \textbf{d}, Close-up of the shielded collisional channel $|{++}\rangle$. At intermediate range the molecules align with respect to the intermolecular axis.}
\label{fig:setup}
\end{figure}

In our experiment, the microwave field is generated by a helical antenna, as illustrated in Fig.~\ref{fig:setup}a. The antenna emits a mainly $\sigma^{-}$-polarized microwave that couples the rotational ground state $|J = 0, m_J = 0, \delta N_\text{p} = 0\rangle$ to the excited state $|1, -1, -1\rangle$, as shown in Fig.~\ref{fig:setup}b. Here, $J$ is the rotational quantum number, $m_J$ is its projection on the magnetic field axis, and $\delta N_\text{p}$ is the change in the number of photons in the microwave field. The frequency of the microwave is approximately given by the rotational constant $B_\text{rot} = h \times 2.822$\,GHz \cite{Will2016} and the detuning from resonance $\Delta$ as $2B_\text{rot}/h + \Delta/(2\pi)$. Coupling the rotational states creates the mixed dressed states $|{+}\rangle$ and $|{-}\rangle$. Rotationally excited states with $m_J =0,\,1$, to which the microwave does not couple, remain as spectator states $|{0}\rangle$. On resonance, i.e.\ at $\Delta = 0$, the coupling strength $\hbar\Omega \approx h \times 11$\,MHz defines the splitting of $|{+}\rangle$ and $|{-}\rangle$. We typically choose a blue detuning $\Delta$ on the order of $\Omega$ to prepare the molecules in the $|{+}\rangle$ state (for details on the microwave transition, the coupling strength, and the dressed-state preparation see Supplementary Information).

Molecules that interact in the upper dressed state via their induced rotating dipoles are described by the collisional channel $|{++}\rangle$, shown in Fig.~\ref{fig:setup}c and \ref{fig:setup}d. At long range, their time-averaged interaction energy is given by \cite{Karman2021}
\begin{equation}
V_\text{dd} = -\frac{d_0^2}{4\pi\varepsilon_0 }\frac{1-3\cos^2{\theta}}{12(1+(\Delta/\Omega)^2)R^3},
\end{equation}
where $d_0 \approx 2.7$\,Debye is the intrinsic dipole moment of NaK, $R$ is the intermolecular distance and $\theta$ denotes the angle between the rotation axis of the induced dipole and the intermolecular axis. Molecules thus acquire an effective dipole moment $d_\text{eff} = d_0/\sqrt{12(1+(\Delta/\Omega)^2)}$. Note that the sign of the interaction energy is inverted compared to the interaction between two regular dipoles with dipole moment $d_\text{eff}$.

At intermediate range, the dipole-dipole interaction dominates, so that the molecules orient themselves with respect to the intermolecular axis, rather than along the rotating electric field \cite{Yan2020}. The reorientation is made possible by contributions of the spectator states $|0\rangle$. Since the molecules are prepared in the upper dressed state, they couple at this point to the repulsive branch of the dipole-dipole interaction, which shields the molecules from reaching short range regardless of their initial angle of approach \cite{Karman2019}. In fact, the remaining two-body loss is dominated by non-adiabatic transitions to lower-lying field-dressed states, such as $|{+}0\rangle$, which is accompanied by an energy release on the order of $\hbar \Omega$ and suffices to eject molecules from the optical dipole trap.

\begin{figure}
\centering
\includegraphics[width = \linewidth]{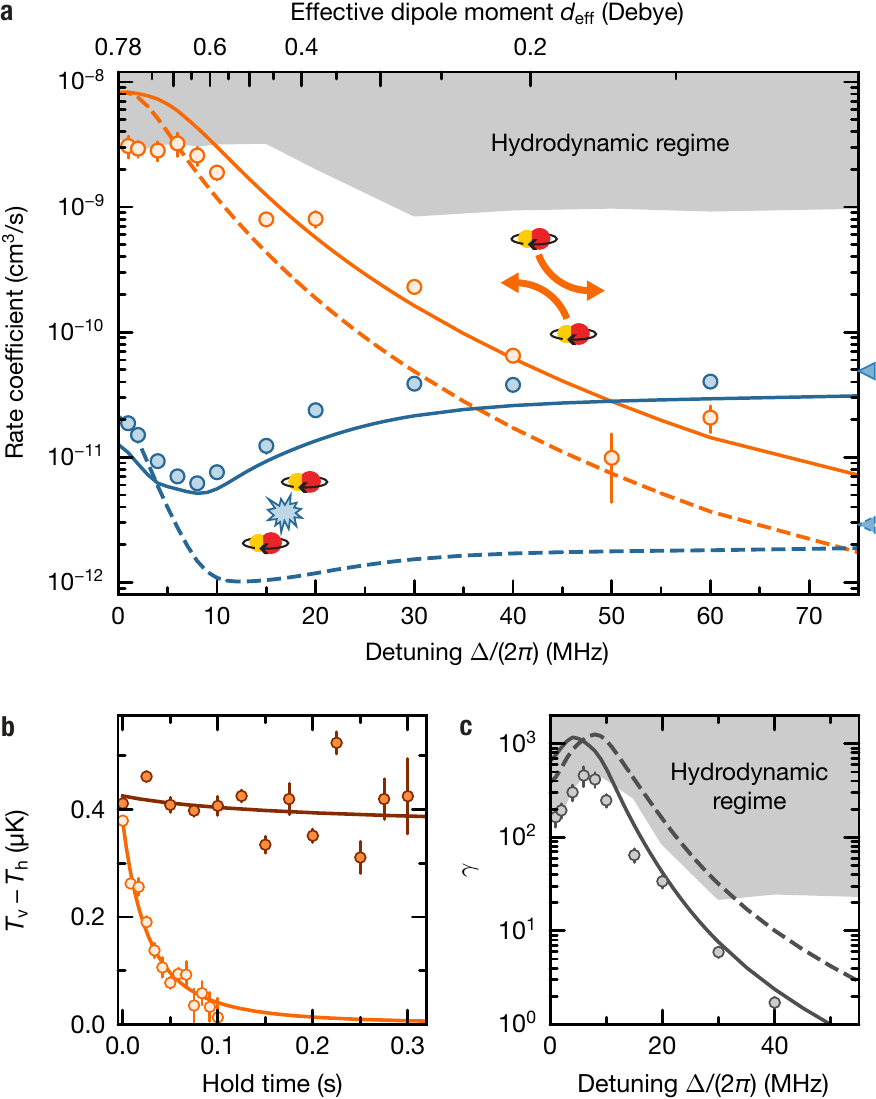}
\caption{\textbf{Elastic and inelastic collisions.} \textbf{a}, Rate coefficients for elastic (orange) and inelastic (blue) scattering events. The solid and dashed lines show coupled channel calculations for a thermal sample with $T=800$\,nK and a degenerate sample with $T=30\,\text{nK} = 0.4\,T_\text{F}$, respectively. The data points show the measurement results for $T=800$\,nK. The shaded area indicates the limit for measurements of the elastic collision rate imposed by the hydrodynamic regime. On resonance, the Rabi frequency is $\Omega \approx 2\pi \times 11$\,MHz. The triangular markers on the side indicate the calculated inelastic rate coefficients in absence of a microwave field. The error bars are the standard deviation from the fit to the differential equations. \textbf{b}, Examples of cross-thermalization measurements at $\Delta = 2\pi \times 30$\,MHz (bright) and $\Delta = 2\pi \times 80$\,MHz (dark). In the latter case, $\beta_\text{el}$ is consistent with zero within the error bars. The error bars are the standard error of the mean of 4 repetitions. The lines are fits of a coupled differential equation system modelling the collision rates (Supplementary Information). \textbf{c}, The ratio $\gamma$ of elastic to inelastic collision rates based on the measurements and calculations presented in \textbf{a}.}
\label{fig:rates}
\end{figure}

The performance of evaporative cooling is ultimately limited by the ratio $\gamma$ of elastic to inelastic two-body collision rates. We characterize the inelastic collision rate coefficient $\beta_\text{in}$ by measuring the two-body decay of the molecules in a thermal gas at temperatures $T$ around 800 nK. The molecules are initially formed from ultracold atoms by means of a magnetic Feshbach resonance and subsequent stimulated Raman adiabatic passage (STIRAP) to their absolute ground state \cite{Bause2021b} (see Supplementary Information for more details on the experimental conditions and the preparation of the molecular samples). For most measurements the initial average density $n_0$ is about $3.0\times10^{11}\,\text{cm}^{-3}$. The two-body loss is relatively high for hot and dense molecules, which helps us to distinguish it from one-body loss caused by phase noise of the microwave (Supplementary Information). We determine the elastic collision rate coefficient $\beta_\text{el}$ by applying parametric heating to the molecular sample along the vertical direction and measuring the cross-dimensional thermalization of effective temperatures $T_\text{h}$ and $T_\text{v}$, as shown in Fig.~\ref{fig:rates}b (see also Supplementary Information). $T_\text{h}$ and $T_\text{v}$ are defined along the horizontal ($x$ and $y$) and the vertical ($z$) directions, respectively. The rate coefficients $\beta_\text{el}$ and $\beta_\text{in}$ are extracted from the time evolution of the measured molecule number $N$ and of the temperatures $T_\text{h}$ and $T_\text{v}$ by fitting a set of coupled differential equations that model the molecule losses and the cross-dimensional rethermalization (Supplementary Information). The results of the measurements are compared with coupled-channel calculations for a thermal sample at $T = 800$\,nK and for a degenerate sample at $30\,\text{nK} = 0.4\,T_\text{F}$, as shown in Fig.~\ref{fig:rates}. The calculations account for a residual ellipticity of the microwave polarization (Supplementary Information).

In absence of the microwave field, we find $\beta_\text{in} = 7.7(5) \times 10^{-11}\,\text{cm}^3/\text{s}$ for $T=800$\,nK, which is in reasonable agreement with the calculated value of $4.9 \times 10^{-11}\,\text{cm}^3/\text{s}$. The shielding is most efficient at $\Delta = 2\pi \times 8$\,MHz, where $\beta_\text{in}$ drops to $6.2(4) \times 10^{-12}\,\text{cm}^3/\text{s}$, corresponding to an order of magnitude suppression of the two-body losses, as illustrated in Fig.~\ref{fig:rates}a. 

For spin-polarized fermionic polar molecules, the elastic collision rate is dominated by dipolar scattering. The corresponding scattering cross section scales approximately as $d_\text{eff}^4$ (Supplementary Information) and can therefore be tuned over multiple orders of magnitude with the detuning of the microwave. In the regime of weak interactions, i.e., at large detunings, the rethermalization rate is proportional to $\beta_\text{el}$. However, for $\Delta\leq 2\pi\times10$\,MHz the elastic collisions become so frequent that the mean free path of the molecules is less than the size of the molecular cloud, even though here, we intentionally reduced the initial density to $n_0=0.7\times10^{11}\,\text{cm}^{-3}$. In this hydrodynamic regime, the rethermalization rate is limited to about $\bar{\omega}/(2\pi) \approx 120$\,Hz, where $\bar{\omega}$ is the geometric mean trap frequency \cite{Ma2003}. Consequently, measured values of $\beta_\text{el}$ saturate near the so-called hydrodynamic limit $N_\text{col}\bar{\omega}/(2\pi n_0)$ where $N_\text{col}\approx 2$ is the average number of collisions required for rethermalization in our system (Supplementary Information). This also limits the maximal measured value of $\gamma$ to 460(110), as illustrated in Fig.~\ref{fig:rates}c. Away from the hydrodynamic regime, we find excellent agreement between the experimentally determined and the calculated values of $\beta_\text{el}$ and $\gamma$ for $T=800$\,nK. Our calculations show that $\gamma$ can exceed $1000$ for ideal values of $\Delta$. In the future, it should be possible to improve the shielding and achieve $\gamma \approx 5000$ by optimizing the purity of the microwave polarization (Supplementary Information).

\section*{Evaporative cooling to a deeply degenerate Fermi gas of molecules}
With $\gamma\gtrsim500$ at the optimum shielding detuning $\Delta=2\pi\times 8$\,MHz, the evaporative cooling of our molecular sample is straightforward. We start with a low-entropy but non-thermalized sample of about $2.5\times 10^4$ molecules produced from a density-matched degenerate atomic mixture \cite{Duda2021}. The power of the two horizontally propagating laser beams, which hold the molecules against gravity and thereby define the effective trap depth $U_\text{trap}$, is lowered exponentially over the course of $150$\,ms. The hottest molecules can then escape the dipole trap in the direction of gravity. The remaining molecules rethermalize via elastic dipolar collisions, effectively reducing $T$ and $T/T_\text{F}$. During evaporation, the calculated elastic collision rate is about 500 Hz, much higher than the trap frequencies. Therefore, the rethermalization rate saturates to around $\bar{\omega}/(2\pi)\approx 60$\,Hz. To maintain a high rethermalization rate while lowering the trap depth, we reinforce the horizontal confinement by exponentially ramping up the power of an additional beam propagating along the vertical direction.

\begin{figure}
\centering
\includegraphics[width = \linewidth]{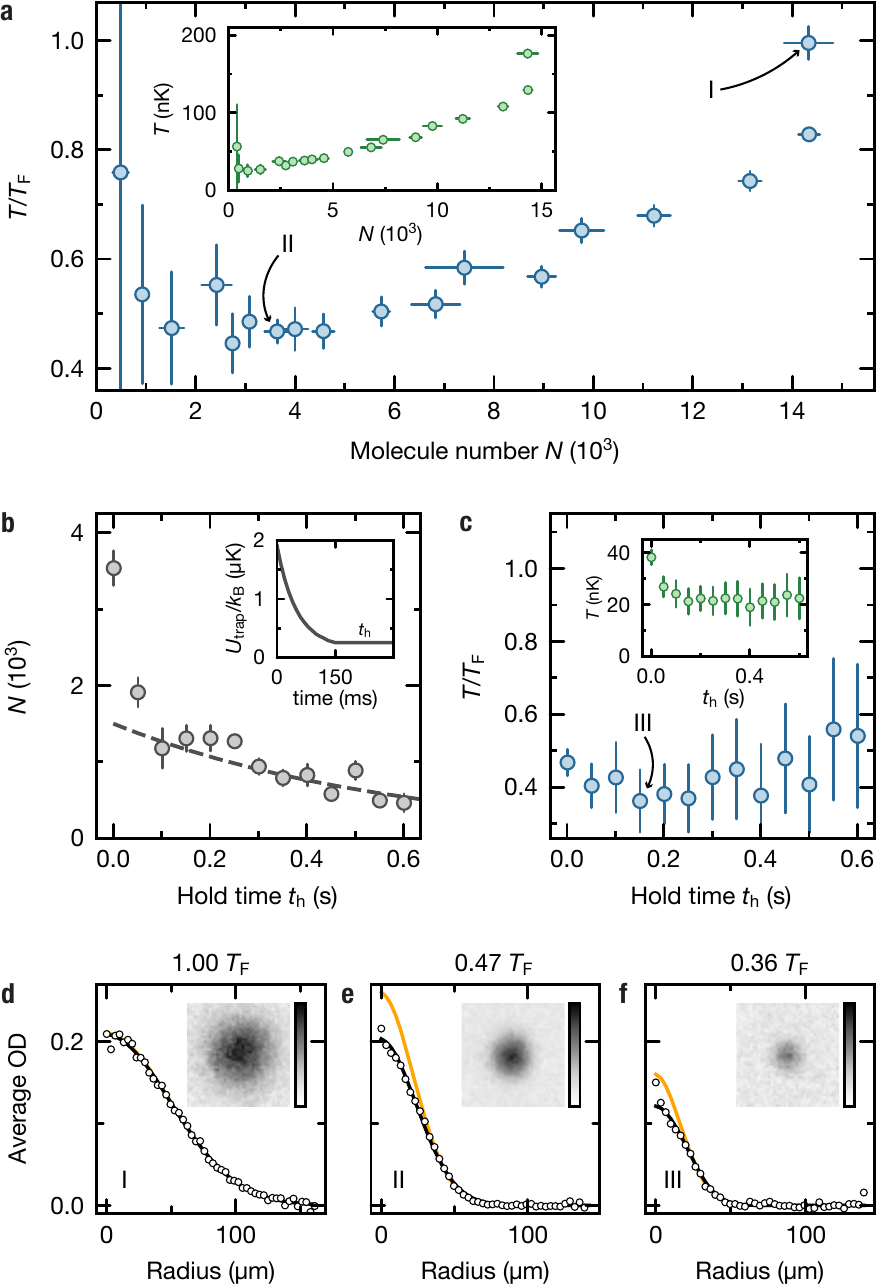}
\caption{\textbf{Evaporation.} \textbf{a}, $T/T_\text{F}$ and $T$ (inset) against the remaining number of molecules $N$ after $150$\,ms evaporative cooling for various final trap depths. \textbf{b}, Plain evaporation during the hold time $t_\text{h}$ after $150$\,ms of forced evaporation. The dashed line indicates the one-body decay and the inset sketches the evolution of the effective trap depth $U_\text{trap}$. \textbf{c},  $T/T_\text{F}$ and $T$ (inset) during the hold time $t_\text{h}$. The error bars of $N$ in \textbf{a} and \textbf{b} are the standard error of the mean of 5--20 repetitions. The error bars of $T$ in \textbf{a} and \textbf{c} are the standard deviation from the fit to the averaged images. \textbf{d--f}, Azimuthally averaged optical density of the samples after $10$\,ms time of flight. The samples are prepared under the evaporation conditions I \textbf{(d)}, II \textbf{(e)}, and III \textbf{(f)}. The images of the samples are averages from $5$ (\textbf{d}), $20$ (\textbf{e}) or $6$ individual images (\textbf{f}). The black lines show polylogarithmic fit functions, while the orange lines are fits of a Gaussian to the thermal wings of the sample.}
\label{fig:evap}
\end{figure}

We characterize the evaporation by varying the final trap depth. Figure~\ref{fig:evap}a shows $T$ and $T/T_\text{F}$ against the number of remaining molecules $N$ after 150 ms forced evaporation. The values of $T$ and $T/T_\text{F}$ are deduced from a polylogarithmic fit to the momentum distribution of the sample, which is imaged after $10$\,ms time of flight (Supplementary Information). If the trap depth is not reduced, the interacting molecules will thermalize but are not forced to evaporate. Initially the molecules exhibit a sloshing motion in the trap due to photon-recoil transfer from the STIRAP pulses. This motion damps out while the molecules thermalize, which effectively reduces the phase space density of the sample. After $150$\,ms holding at the initial trap depth, we obtain $1.43(5)\times 10^4$ molecules at a temperature of $176(5)$\,nK with $T/T_\text{F} = 1.00(3)$ (Fig.~\ref{fig:evap}d). If we instead evaporate down to $3.6(3)\times 10^3$ molecules by reducing the final trap depth to about $k_\text{B}\times250$\,nK, where $k_\text{B}$ is the Boltzmann constant, we reach a temperature of $38(2)$\,nK with $T/T_\text{F} = 0.47(2)$ (Fig.~\ref{fig:evap}e). However, the evaporation has not stopped at this point. We can hold the molecular sample for an additional hold time $t_\text{h}$ in the trap to make use of plain evaporation, as shown in Fig.~\ref{fig:evap}b and \ref{fig:evap}c. The plain evaporation lasts for about $100$\,ms. Thereafter, the loss of molecules is dominated by one-body loss caused by phase noise of the microwave, which results in a $1/e$ lifetime of $590(100)$\,ms (Supplementary Information). At $t_\text{h} = 150$\,ms, we measure a record low temperature of $21(5)$\,nK with $T/T_\text{F} = 0.36(9)$ (Fig.~\ref{fig:evap}f).

We checked that the temperatures obtained from the Fermi--Dirac fit are consistent with those deduced from a Gaussian fit to the thermal wing, in which a possible influence of the dipolar interactions is relatively small (Supplementary Information). Interestingly, the optical density in a small region of the cloud center is higher than the Fermi--Dirac fit in the coldest sample III. However, the low signal-to-noise ratio does not allow us to conclusively establish whether this is due to imaging noise or potential dipolar interaction effects. In the future, it will be interesting to investigate the underlying mechanism with improved evaporation and detection.

\section*{Discussion}
Our highest measured value of $\gamma$ of about $500$ is a conservative estimate due to the limitations of our measurement method in the hydrodynamic regime. Our numerical models, which agree well with the data outside of the hydrodynamic regime, predict a ratio of $\gamma \geq 1000$. This is almost a factor of 100 higher than the ratio realized in previous experiments in 3D \cite{Li2021} and almost an order of magnitude higher compared to experiments in 2D \cite{Valtolina2020}. Such optimal collisional parameters facilitate the efficient evaporation of NaK molecules to below 0.4 $T_\text{F}$ compared to previous evaporation results reaching 1.4 $T_\text{F}$ in 3D \cite{Li2021} and 0.6 $T_\text{F}$ in 2D \cite{Valtolina2020}. 

With the coldest samples realized in our experiment, the dipolar interaction in the system corresponds to about $5$\% of the Fermi energy. This is three times higher than what was reached in degenerate Fermi gases of magnetic atoms \cite{Aikawa2014}. In the near future, intriguing dipolar many-body phenomena such as modifications of collective excitation modes \cite{Wachtler2017}, distortion \cite{Aikawa2014} or the collapse \cite{Veljic2019} of the Fermi sea should be observable in suitable trap geometries and with improved detection of the cloud expansion.

The reduced temperature $T/T_\text{F}$ reached here is only a factor of about four higher than the predicted critical temperature of topological $p$-wave superfluidity with strong dipolar interactions \cite{Baranov2002,Cooper2009,Shi2010}. 
The performance of the evaporation can be further improved by implementing the following measures: first, the initial phase space density can be increased by optimizing the STIRAP transfer. Second, lower phase noise and better polarization purity of the microwave field should lead to reduced one- and two-body losses during evaporation. Finally, new strategies of evaporation in the hydrodynamic regime need to be explored to accelerate the thermalization process \cite{Ma2003}. 
With these upgrades, it should be possible to reach temperatures below 0.1 $T_\text{F}$, where many intriguing quantum phases are expected \cite{Baranov2002,Cooper2009,Shi2010,Schmidt2021,Wu2016,Bruun2008,Matveeva2012,Baranov2012}. Especially, fermionic polar molecules can pair up and form a superfluid with anisotropic order parameter and even a Bose--Einstein condensate of tetramers \cite{Baranov2002,Shi2010}.
Up to now, such scenarios have rarely been theoretically investigated because it was believed that polar molecules could not be sufficiently stable under conditions where both attractive and repulsive interactions play an important role.

\section*{Conclusion}
In conclusion, we have demonstrated a general and efficient approach to evaporatively cool ultracold polar molecules to deep quantum degeneracy in 3D by dressing the molecules with a blue-detuned circularly polarized microwave, achieving record low temperatures together with strong tunable dipolar interactions. The simplicity of the technical setup makes our method directly applicable in a wide range of ultracold molecule experiments. Our results point to an exciting future of long-lived degenerate polar molecules for investigating novel quantum many-body phases with long-range anisotropic interactions and for other applications in quantum sciences.

\section*{Acknowledgements}
We thank Y. Bao, L. Anderegg, A. Pelster and A. Bala\v{z}, and T. Shi for stimulating discussions, B. Braumandl for the simulation of the microwave field, F. Deppe for lending the ultra low-noise microwave signal generator, and T. Hilker for critical reading of the manuscript. We gratefully acknowledge support from the Max Planck Society, the European Union (PASQuanS Grant No.\ 817482) and the Deutsche Forschungsgemeinschaft under Germany's Excellence Strategy -- EXC-2111 -- 390814868 and under Grant No.\ FOR 2247. A.S.\ acknowledges funding from the Max Planck Harvard Research Center for Quantum Optics.

\clearpage

\renewcommand{\figurename}{Fig.}
\renewcommand\thefigure{S\arabic{figure}}    
\setcounter{figure}{0}
\renewcommand{\theequation}{S\arabic{equation}}
\setcounter{equation}{0}

\section*{Supplementary Information}

\subsection*{Sample preparation}
To create our molecular samples, we first prepare a density-matched double-degenerate mixture of $^{23}$Na and $^{40}$K atoms. The atoms are subsequently associated to weakly bound molecules by means of a magnetic Feshbach resonance. Finally, the molecules are transferred to their absolute ground-state via STIRAP. Details about the preparation process are described in Refs.~\cite{Duda2021, Bause2021b}. At the beginning of the measurements described in the main text, the molecules are trapped by the 1064-nm and the 1550-nm beam shown in Fig.~\ref{fig:setup}a at a dc magnetic field of $72.35$\,G.

For the measurements of the collision rates, the microwave transition strength, and to characterize the one-body loss, we work with thermal molecules and sometimes reduce the molecule number to suppress interactions. For the collision rate measurements, the trap frequencies are $(\omega_x, \omega_y, \omega_z) = 2\pi \times (67,99,244)$\,Hz. For the evaporation, on the other hand, we start with near degenerate molecules at $2\pi \times (45,67,157)$\,Hz and end up, e.g., at $2\pi \times (52,72,157)$\,Hz in case I or at $2\pi \times (42,56,99)$\,Hz in case II and III (see Fig.~\ref{fig:evap}).

To measure the cross-dimensional thermalization, we heat the weakly bound molecules along the vertical direction after we separated them from unbound atoms and before STIRAP is applied. For this purpose, we use parametric heating by modulating the intensity of the 1064-nm beam at twice the vertical trap frequency.

\subsection*{Microwave field generation}
It is essential that the phase noise of the microwave source does not induce transitions between the dressed states. We generate the microwave with a vector signal generator (Keysight E8267D). The microwave passes through a voltage controlled attenuator (General Microwave D1954) before it is amplified with a $10$-W power amplifier (KUHNE electronic KU PA 510590 -- 10 A). At $10$\,MHz carrier offset we measure $-150\,\text{dBc}/\text{Hz}$ phase noise density from the signal generator and no significant enhancement from the amplifier. The microwave is emitted by a $5$-turn helical antenna (customized by Causemann Flugmodellbau) whose top end is about $2.2$\,cm away from the molecular sample.

With the voltage controlled attenuator we can adiabatically prepare the molecules in the dressed state by ramping the power attenuation linearly within $100$\,µs over a range of $65$\,dB.

\subsection*{Imaging and thermometry}
To image the molecules, we transfer them back into the non-dressed absolute ground-state by ramping down the microwave power. Subsequently, the dipole traps are turned off and return STIRAP pulses are applied to bring the molecules back into the weakly bound state. After time of flight, typically $10$\,ms, the atoms are dissociated by ramping the magnetic field back over the Feshbach resonance. The magnetic field has to cross the Feshbach resonance slowly to minimize the release energy. In the end, the dissociated molecules are imaged via absorption imaging. We estimate that the derived temperature of the molecular sample could be overestimated by about $7$\,nK due to the residual release energy. Note that the values of $T$ and $T/T_\text{F}$ reported in the main text do not account for the release energy.

To obtain the temperature of the molecular sample, we fit the absorption images with the Fermi--Dirac distribution
\begin{equation}
n_{\mathrm{FD}}(x,z) = n_{\mathrm{FD,0}}\, \mathrm{Li}_{2} \left [- \zeta \exp(-\frac{x^2}{2 \sigma_x^2}-\frac{z^2}{2 \sigma_z^2})\right ],
\end{equation}
where $n_{\text{FD},0}$ is the peak density, $\mathrm{Li}_{2}(x)$ is the dilogarithmic function, $\zeta$ is the fugacity and $\sigma_{i={x,z}}$ are the cloud widths in the $x$- and $z$-direction. Given a cloud width $\sigma_{i}$, we can calculate the temperature $T_{i}$ by
\begin{equation}
\sigma_{i} = \frac{\sqrt{1+\omega_{i}^2 t_\text{TOF}^2}}{\omega_{i}}\sqrt{\frac{k_\text{B} T_{i}}{m}},
\end{equation}
where $\omega_{i}$ is the trapping frequency in the $i$-th direction, $t_\text{TOF}$ is the time of flight and $m$ is the mass of the molecules. The fugacity can be associated to the ratio of the temperature $T$ and the Fermi temperature $T_\text{F}$ with the relation
\begin{equation}
\left ( \frac{T}{T_\text{F}} \right)^3 = -\frac{1}{6\,\mathrm{Li}_3(-\zeta)},
\end{equation}
where $\mathrm{Li}_3(x)$ is the trilogarithmic function. $T_\text{F}=(6N)^{1/3} \hbar \bar{\omega} /k_\text{B}$ is given by the molecule number $N$ and the geometric mean trap frequency $\bar{\omega}=(\omega_x \omega_y \omega_z)^{1/3}$. By rewriting $\zeta$ and fixing $T_\text{F}$, we are only left with the fitting parameters $n_{\text{FD},0}$, $T_x$, and $T_z$. We note that the temperature in the direction of the imaging beam $T_y$ is assumed to be equal to $T_x = T_\text{h}$.

In addition, we independently determine the temperatures of the molecular samples from the time-of-flight images by fitting the thermal wings of the cloud to a Gaussian distribution
\begin{equation}
n_{\mathrm{th}}(x,z) = n_{\text{th},0} \exp(-\frac{x^2}{2 \sigma_x^2}-\frac{z^2}{2 \sigma_z^2}),
\end{equation}
where $n_{\text{th},0}$ is the peak density. Similar to Ref.~\cite{Valtolina2020}, we first fit a Gaussian distribution to the whole cloud. We then constrain the Gaussian distribution to the thermal wings of the cloud by excluding a region of $1.5 \sigma$ around the center of the image. We find that by excluding $1.5 \sigma$, the ratio of signal to noise allows for the fit to converge for all datasets in Fig.~\ref{fig:evap}a.

\begin{figure}
\centering
\includegraphics[width = \linewidth]{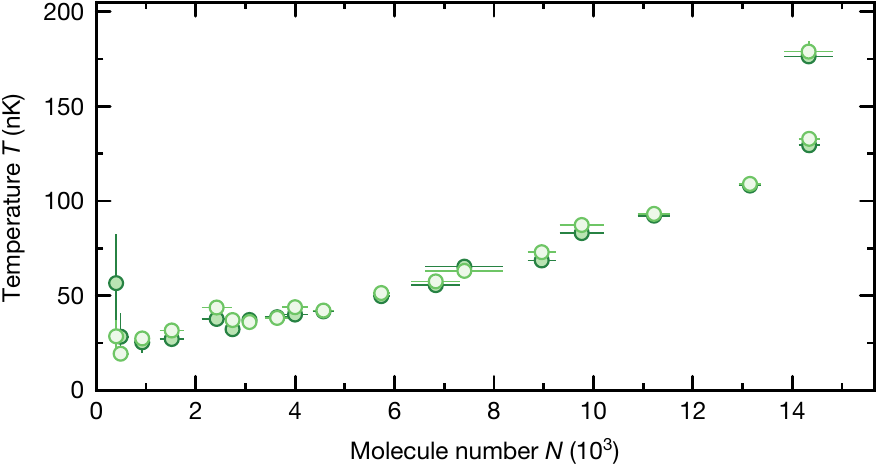}
\caption{\textbf{Thermometry.} Temperatures extracted from a Fermi--Dirac distribution (dark) and a Gaussian distribution to the thermal wings (bright) against the number of molecules $N$ after $150$\,ms evaporative cooling for various final trap depths. The temperatures are extracted from averaged images. The error bars of $N$ are the standard error of the mean of 5--20 repetitions. The error bars in $T$ are the standard deviation from the fit.}
\label{fig:thermometry}
\end{figure}

The temperatures extracted from fitting the Fermi--Dirac distribution and fitting the Gaussian distribution to the thermal wings are compared in Fig.~\ref{fig:thermometry}.

\subsection*{Model for elastic and inelastic collisions}
The elastic and inelastic collision rate coefficients $\beta_\text{el}$ and $\beta_\text{in}$ are experimentally determined from the time evolution of the measured molecule number $N$, the average temperature $(2T_\text{h} + T_\text{v})/3$ and the differential temperature $T_\text{v}-T_\text{h}$ by numerically solving the differential equations \cite{Ni2010dipolar,Li2021}
\begin{equation}
    \frac{dN}{dt} = \left( -K\frac{2T_\text{h} + T_\text{v}}{3} n - \Gamma_1 \right) N,
\end{equation}
\begin{equation}
    \frac{dT_\text{h}}{dt} = \frac{1}{12} K T_\text{v} T_\text{h} n + \frac{\Gamma_\text{th}}{3} (T_\text{v} - T_\text{h}),
\label{equ:th}
\end{equation}
\begin{equation}
    \frac{dT_\text{v}}{dt} = \frac{1}{12}K (2T_\text{h} - T_\text{v}) T_\text{v} n - 2 \frac{\Gamma_\text{th}}{3} (T_\text{v} - T_\text{h}),
\label{equ:tv}
\end{equation}
with the mean density
\begin{equation}
    n = \frac{N}{8\sqrt{\pi^3 k_\text{B}^3 T_\text{h}^2 T_\text{v} / m^3 \bar\omega^6}}.
\end{equation}
Here, $K$ is the temperature-independent two-body loss coefficient, averaged for simplicity over all collision angles, and
\begin{equation}
    \Gamma_\text{th} = \frac{n \sigma_\text{el}v}{N_\text{col}}
\end{equation}
is the rethermalization rate with the elastic scattering cross section $\sigma_\text{el}$ and the thermally averaged collision velocity
\begin{equation}
    v = \sqrt{16 k_\text{B} (2T_\text{h} + T_\text{v})/(3\pi m)}.
\end{equation}
The average number of collisions per rethermalization is taken from Ref.~\cite{Wang2021} as
\begin{equation}
    N_\text{col} = \bar{\mathcal{N}}_{z}(\phi) = \frac{112}{45 + 4\cos(2\phi) - 17\cos(4\phi)}
\end{equation}
where $\phi$ is the tilt of the dipoles in the trap, which, in our case, corresponds to the tilt of the microwave wave vector with respect to the dc magnetic field. Following our characterization of the microwave polarization, we assume $\bar{\mathcal{N}}_{z}(29^\circ) = 2.05$.

The anti-evaporation terms, i.e., the first terms in Eqs.~\ref{equ:th} and \ref{equ:tv}, assume a linear scaling of the two-body loss rate with temperature. Our calculations predict that this assumption does not hold for small detunings ($\Delta < 2\pi \times 20$\,MHz), as illustrated in Fig.~\ref{fig:rates}. Our results, however, do not significantly change when we instead assume no temperature dependence in this regime.

Finally, after determining $\sigma_\text{el}$ and $K$, the elastic and inelastic collision rate coefficients
\begin{equation}
    \beta_\text{el} = \sigma_\text{el} v
\end{equation}
and
\begin{equation}
    \beta_\text{in} = K(2 T_\text{h} + T_\text{v})/3
\end{equation}
are plotted in Fig.~\ref{fig:rates} assuming a fixed temperature $T = T_\text{h} = T_\text{v}$.

\begin{figure}
\centering
\includegraphics[width = \linewidth]{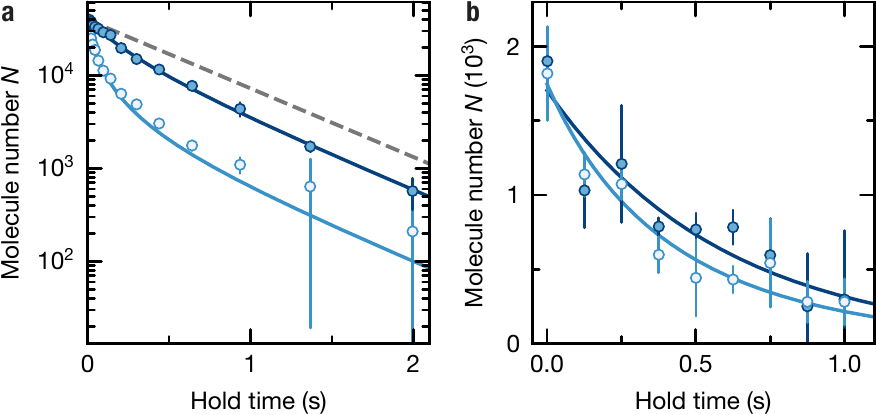}
\caption{\textbf{One- and two-body loss.} \textbf{a}, Molecule loss with (dark) and without (bright) microwave at high initial densities. The gray dashed line shows the one-body contribution. The lines are fits to the differential-equation model. \textbf{b}, Loss at low initial densities from which the one-body loss rate $\Gamma_1$ is determined. The lines are exponential fit functions. In both subfigures, bright blue data are taken without a microwave field, while dark blue data are taken with a microwave field at $\Delta=2\pi\times 8$\,MHz. The error bars are the standard error of the mean of two (\textbf{a}) or three (\textbf{b}) repetitions.}
\label{fig:onebodyloss}
\end{figure}

Example data of the loss measurements, performed to determine $\beta_\text{in}$, are shown in Fig.~\ref{fig:onebodyloss}a. At high densities, two-body loss is the dominating contribution, while at low densities, the exponential shape of the loss curve shows that one-body effects outweigh collisions.
In order to limit the number of free fit parameters, we determine $\Gamma_1 = 1.7(4)$\,Hz in independent measurements at low densities, as shown in Fig.~\ref{fig:onebodyloss}b.
To avoid confounding effects from collisions, we reduce the initial molecule number to about 2000 for these measurements. Under these conditions, the $1/e$ lifetime is $570(100)$ ms without shielding. Turning on the shielding results in a similar $1/e$ lifetime of about $590(100)$ ms. The lifetime reduces by a factor two when a microwave source with a 3 dB higher phase-noise density is used. This observation is consistent with the assumption that the noise power spectral density around $2\pi\times 10$\,MHz offset from the carrier is a limiting factor of the one-body lifetime of the dressed state, even at a level of  $-150\,\text{dBc}/\text{Hz}$ \cite{Anderegg2021}.

\begin{figure}
\centering
\includegraphics[width = \linewidth]{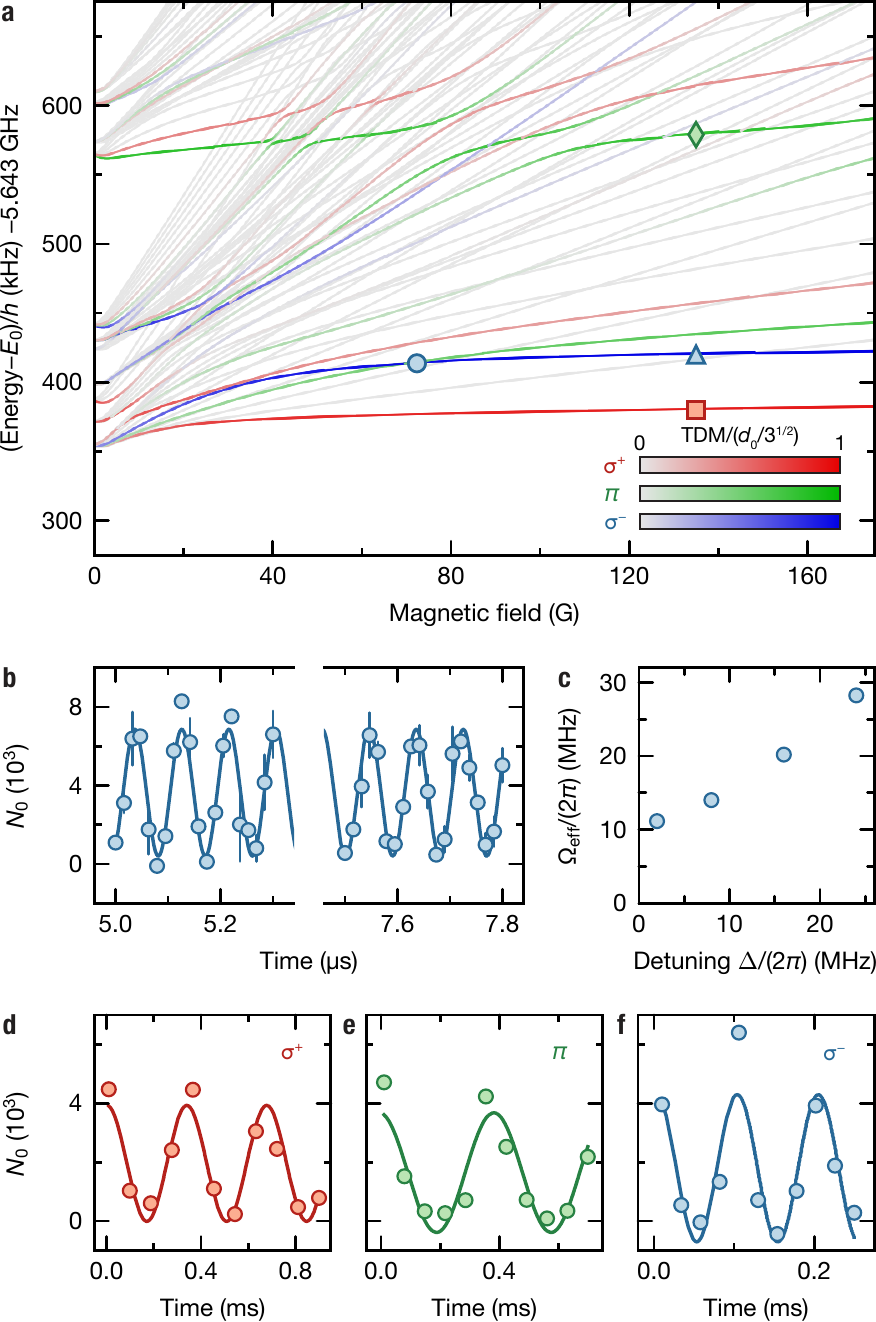}
\caption{\textbf{Microwave transitions.} \textbf{a}, Zeeman diagram of the microwave transition frequencies from the absolute ground state with energy $E_0$ to the hyperfine manifold of the exited rotational state ($J=1$). The color scales show the transition dipole moment (TDM) for the different microwave polarizations, assuming that the ground state has the nuclear spin projections $(m_{i,\text{Na}},m_{i,\text{K}}) = (3/2,-4)$, which is a good approximation for magnetic fields larger than $3$\,G. The state energies and TDMs are calculated with the code from Ref.~\cite{Blackmore2020} and the parameters from Ref.~\cite{Will2016}. The circle marks the $\sigma^{-}$ transition that is used for microwave shielding. The diamond, triangle, and square mark the microwave transitions that are used to probe the polarization of the microwave field. \textbf{b}, Example for off-resonant Rabi oscillations at full microwave power with a detuning of $2\pi \times 2$\,MHz from the $\sigma^{-}$ transition at $72.35$\,G. The data show the number of molecules in the absolute ground state $N_0$ and the error bars are the standard error of the mean of two repetitions. The line is a sinusoidal fit function that yields an effective Rabi frequency $2\pi \times 11.157(6)$\,MHz. \textbf{c}, Effective Rabi frequency $\Omega_\text{eff}$ at full microwave power versus detuning from the $\sigma^{-}$ transition at $72.35$\,G. \textbf{d--f}, Examples of resonant Rabi oscillations at the $\sigma^{+}$ (\textbf{d}), $\pi$ (\textbf{e}), and $\sigma^{-}$ (\textbf{f}) transitions indicated in \textbf{a} at $135$\,G. The sinusoidal fit functions yield Rabi frequencies of $2\pi \times 2.96(8)$\,kHz, $2\pi \times 2.58(13)$\,kHz, and $2\pi \times 9.9(4)$\,kHz, respectively. The power of the microwave is attenuated by $55$\,dB (\textbf{d}) or $61$\,dB (\textbf{e} and \textbf{f}). }
\label{fig:transitions}
\end{figure}

\subsection*{Microwave transitions}
Figure~1 of the main text shows only the most essential parts of the energy level structure. The full hyperfine structure of the excited rotational state $J=1$ is presented in Fig.~\ref{fig:transitions}a. In contrast to the situation in CaF \cite{Anderegg2021}, NaK has no fine structure in the electronic ground state. Consequently the rotationally excited states that can couple to the absolute ground state, which have $(m_{i,\text{Na}},m_{i,\text{K}}) = (3/2,-4)$ character, are spread over just a few hundred kilohertz, much less than the microwave detunings used for shielding. Here, $m_{i,\text{Na}}$ and $m_{i,\text{K}}$ are the projections of the nuclear spins of Na and K onto the magnetic field axis, respectively. At $72.35$\,G a $\sigma^{-}$-polarized microwave couples mainly to the $|J=1,m_J=-1\rangle$ state with transition frequency $5.643\,4137$\,GHz. In absence of electric fields, the transition dipole moment (TDM) is already $0.96\,d_0/\sqrt{3}$. In a strong microwave field, the nuclear spin projections are further purified bringing the TDM even closer to the maximum value of $d_0/\sqrt{3}$.

Unfortunately, we cannot directly measure the coupling strength $\Omega$ at full microwave power by driving resonant Rabi oscillations. Coupling to weaker transitions would lead to beat signals with the main Rabi oscillations. However, we can measure the effective Rabi frequency $\Omega_\text{eff}$ at detunings that are large enough to suppress coupling to unwanted transitions. For these measurements we temporarily switch off the optical dipole traps to avoid ac Stark shifts from the trapping light. We then generate a rectangular microwave pulse with a fast microwave switch that controls the input of our microwave amplifier. The microwave pulse drives Rabi oscillations between the rotational states, as shown in Fig.~\ref{fig:transitions}b. We ignore the oscillations during the first $5$\,µs of the microwave pulse, as the amplifier requires some time to reach full output power. These measurements are performed at low molecule density in order to suppress dephasing and inelastic collisions between the molecules. From the measurements of $\Omega_\text{eff}$ shown in Fig.~\ref{fig:transitions}c we deduce $\Omega \approx 2\pi \times 11$\,MHz.

\subsection*{Microwave polarization}
It is crucial that the microwave field has a high degree of polarization purity in order to achieve efficient shielding. Especially for small detunings, unwanted polarization components can couple to excited states with different $m_J$ character which significantly reduces the shielding effect \cite{Karman2019}. Fortunately, the strong ac electric field of the microwave redefines the quantization axis of the molecules so that we only have to consider $\sigma^{+}$ and $\sigma^{-}$ components in the microwave frame. We can use different microwave transitions of the molecules to characterize the polarization of the microwave field in situ. However, in order to resolve the individual transitions, only weak microwave fields can be applied, i.e., the microwave polarization can only be characterized in the frame of the dc magnetic field. We probe the microwave field polarization at $135$\,G, where we can still stabilize the dc magnetic field and where the used transitions, marked in Fig.~\ref{fig:transitions}a, are reasonably isolated. The measurements, shown in Fig.~\ref{fig:transitions}(d--f), are performed similarly to the measurements of $\Omega_\text{eff}$, described earlier. However, here we measure on resonance and the microwave power is attenuated by $55$--$61$\,dB. The microwave power has to be low enough to avoid off-resonant coupling to neighbouring transitions but strong enough to realize Rabi oscillations of at least $2\pi \times 2$\,kHz, because we can only turn off the dipole traps for about $1$\,ms before we start losing molecules. The TDMs of the selected $\sigma^{+}$, $\pi$, and $\sigma^{-}$ transitions are $0.875\,d_0/\sqrt{3}$, $0.789\,d_0/\sqrt{3}$, and $0.989\,d_0/\sqrt{3}$, respectively. From the measured Rabi frequencies, the relative microwave power, and the TDMs, we can determine the ratio of the electric field amplitudes $E_{\sigma^{+}}/E_{\sigma^{-}} = 0.169(8)$ and $E_{\pi}/E_{\sigma^{-}} = 0.462(30)$. Although we do not know the phase relation between the measured ac electric field components in the frame of the dc magnetic field, we can deduce that the wave vector of the microwave is tilted somewhere between $21.5(12)^\circ$ and $29.0(16)^\circ$ with respect to the magnetic field axis. In the microwave frame the ellipticity angle is then given by the electric field amplitudes $E_{\sigma^{+}}^\prime$ and $E_{\sigma^{-}}^\prime$ as $\xi = \arctan(E_{\sigma^{+}}^\prime/E_{\sigma^{-}}^\prime)$ and has a value between $11.5(5)^\circ$ and $5.9(6)^\circ$. To calculate the potential curves in Fig.~1 of the main text and the rate coefficients in Fig.~2a of the main text we assume $\xi = 6.2^\circ$.

\begin{figure}
\centering
\includegraphics[width = \linewidth]{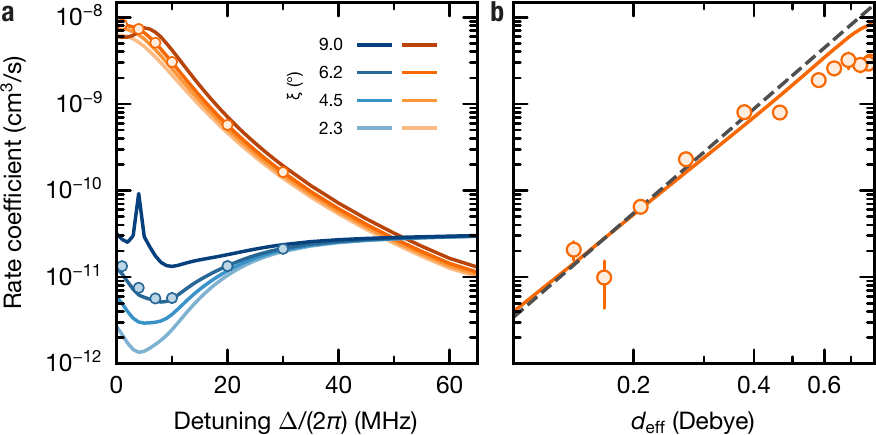}
\caption{\textbf{Scaling of the collision rates.} \textbf{a}, Calculations of the elastic (orange) and inelastic (blue) collision rate coefficients for various microwave ellipticity angles $\xi$. Further simulation parameters are $T=800$\,nK and $\Omega = 2\pi \times 11$\,MHz. The markers are calculations including hyperfine interactions for $\xi = 6.2^\circ$.  \textbf{b}, The circles and the solid line are the measured values and the calculations of the elastic rate coefficients, as shown in Fig.~2 of the main text for $T=800$\,nK. The dashed line is a simplified calculation of the elastic scattering rate based on the effective dipole moment $d_\text{eff}$. The error bars are the standard deviation from the fit to the differential equations.}
\label{fig:ellipticity}
\end{figure}

The ellipticity of the microwave polarization has a significant effect on the inelastic collision rate and therefore on the shielding efficiency, as illustrated in Fig.~\ref{fig:ellipticity}a. For large enough ellipticity even inelastic scattering resonances can arise when the undesired circular-polarization component contributes to the coupling at small detunings. In our setup the purity of the circular polarization can partially be optimized by moving and rotating a cylindrical metal sheet that surrounds the helical antenna.

\subsection*{Coupled-channels calculations}
We perform coupled-channels scattering calculations using the framework developed in Refs.~\cite{Karman2018,Karman2019,karman2020microwave} and here we summarize numerical details of these calculations.

The Hamiltonian describes the NaK molecules as rigid rotors with electric dipole moments that interact with one another as well as with the microwave electric field.
Furthermore, the molecules have nuclear spins that couple with one another and with a static magnetic field.
The channel basis was truncated by including only the lowest two rotational states $J=0,1$ and partial waves $L=1,3,5$.
We propagated the scattering wavefunctions from $R_\mathrm{min}=30\,a_0$ to $R_\mathrm{max}=60\,000\,a_0$,
imposing a capture boundary condition at $R_\mathrm{min}$ and the usual scattering boundary conditions at $R_\mathrm{max}$, from which the $S$-matrix and collision cross sections are obtained.
This short-range boundary results in loss rates given by the universal loss model \cite{Idziaszek2010universal} in the absence of external fields,
which is in reasonable but not perfect agreement with experimental loss rates, $4.9 \times 10^{-11}\,\text{cm}^3/\text{s}$ versus $7.7(5) \times 10^{-11}\,\text{cm}^3/\text{s}$ at $T=800$\,nK, respectively.
We performed scattering calculations for nine values of the collision energy spaced logarithmically between $0.1\,k_\text{B}T$ and $10\,k_\text{B}T$, and subsequently cross sections are multiplied by the velocity and averaged over the Maxwell--Boltzmann or Fermi--Dirac distribution to obtain collision rates compared to experiment in Fig.~2 of the main text.

Scattering rates presented in Fig.~2 of the main text were obtained neglecting hyperfine interactions, and using the microwave polarization determined from the experiment, which is elliptical and tilted with respect to the magnetic field.
We have also performed calculations including hyperfine interactions, but their effect is small, as can be seen in Fig.~\ref{fig:ellipticity}a.
Here we truncated the hyperfine basis by including functions with $\Delta m_i = \pm 1$ only, i.e.,\ functions that differed at most one quantum from the initial state.
Convergence tests with $\Delta m_i = \pm 2$ were also performed.
This figure also contains scattering rates for microwave polarizations of varying ellipticity.
In this case, the polarization lies in the plane perpendicular to the magnetic field axis.

Potential energy curves shown in Fig.~1 of the main text were obtained by diagonalizing the Hamiltonian excluding kinetic energy for fixed $\theta$, the angle between the direction of approach of the colliding molecules and the microwave propagation direction, and for fixed $R$, the distance between the two molecules, as described in Ref.~\cite{Karman2019}.
This omits the centrifugal kinetic energy, which is not well defined for fixed $\theta$,
and it has neglected hyperfine interactions for clarity of Fig.~1 of the main text.

The elastic cross section can, to a reasonable degree, be approximated by $(32\pi/15) (\mu d_\text{eff}^2/(\hbar^2 4\pi\varepsilon_0))^2$ with the reduced mass $\mu = m/2$ \cite{Bohn2009}. A comparison with our coupled-channel calculation is shown in Fig.~\ref{fig:ellipticity}b.

\end{document}